\documentclass[aps,prb,twocolumn,showpacs]{revtex4}
\usepackage{graphicx} 
\def\gapx{\lower 2pt \hbox{$\buildrel>\over{\scriptstyle{\sim}}$\ }}
\def\lapx{\lower 2pt \hbox{$\buildrel<\over{\scriptstyle{\sim}}$\ }}
\def\he4{$^4$He}
\def\paraH2{{\it p}-H$_2$}
\def\Am2{\AA$^{-2}$}

\begin{document}

\widetext
\title{Low temperature phase diagram of condensed {\it para}-Hydrogen in two dimensions}
\author{Massimo Boninsegni} 
\affiliation{Department of Physics, University of Alberta, Edmonton, Alberta, Canada T6G 2J1}
\date{\today}

\begin{abstract}
Extensive Path Integral Monte Carlo simulations of condensed {\it para}-Hydrogen in two dimensions at low temperature have been carried out. In the zero temperature limit, the system is a crystal at equilibrium, with a triangular lattice structure. No metastable liquid phase is observed, as the system remains a solid down to the spinodal density, and breaks down into solid clusters at lower densities.
The equilibrium crystal is found to melt at a temperature close to 7 K.
\end{abstract}
\maketitle
A fluid of {\it para}-Hydrogen (\paraH2)  molecules has long been regarded  as a potential superfluid, due to the light mass and the bosonic character of its constituents \cite{ginzburg72}. In bulk \paraH2,  however, superfluidity (SF) is not observed, because, unlike Helium,  molecular Hydrogen solidifies at a temperature ($T$ $\approx$ 14 K) significantly higher than that ($\sim$ 4 K) at which phenomena such as Bose Condensation and, possibly, SF might occur. This is due the depth of the attractive well of the potential between two Hydrogen molecules, significantly greater  than that between two Helium atoms. Several, attempts  have been made   
\cite{bretz81,maris86,maris87,schindler96} to supercool bulk liquid \paraH2, but no observation of 
SF in the bulk phase of \paraH2 has been reported to date.

Reduction of dimensionality is regarded as a plausible avenue to the stabilization of  a 
liquid phase of \paraH2 at temperatures sufficiently low that a superfluid  transition may be observed.  This
has been the primary motivation underlying the experimental investigation of adsorbed films of \paraH2 on different substrates. For example, the phase diagram and structure of monolayer 
\paraH2 films adsorbed on graphite have been studied by various techniques.
\cite {nielsen80,lauter90,wiechert91,vilches92}
One of the most remarkable aspects is that the melting 
temperature $T_m$ of a solid \paraH2 monolayer can be significantly less than 
bulk \paraH2.  \cite{vilches92} This motivates our interest in the study of the phase diagram of \paraH2 in two mathematical dimensions, which is still relatively unexplored (a systematic study of the one-dimensional phase diagram at zero temperature has been recently carried out \cite{boronat00}).

Some information has been provided in previous numerical work by Gordillo and Ceperley \cite{gordillo97}, and by Wagner and Ceperley; \cite{wagner94,wagner96} for example, it is known that, in the low temperature  limit, the equilibrium phase of the system is a triangular solid, with a two dimensional (2D) density $\theta_\circ$ $\approx$ 0.067 \Am2. On the other hand, little is known about the possible existence of a {\it metastable} liquid phase, at densities below $\theta_\circ$ (i.e., at negative pressure), which one may be able to investigate experimentally by ``stretching" the equilibrium uniform solid phase. Conceivably, such a liquid phase ought to turn superfluid at sufficiently low temperature.

Microscopic calculations for condensed \paraH2 have recently focused on realistic models of adsorbed films, both on  graphite\cite{nho02,nho03}, as well as on alkali metal substrates. \cite{shi03,boninsegni04} 
Here, we present results of a theoretical study of the phase diagram of condensed \paraH2 in 2D, based on path integral Monte Carlo (PIMC) simulations.   The  temperature range explored is between 1 and 8 K; we extrapolate the results obtained at low temperature to obtain the $T=0$ thermodynamic equation of state. 

In agreement with previous calculations, we find that the equilibrium phase of the system at $T$=0 is a triangular crystal; we estimate the equilibrium density $\theta_\circ$=0.0668$\pm$0.0005 \Am2. We also estimate the {\it spinodal} density $\theta_s$, namely the lowest density down to which the uniform phase can be stretched, before becoming unstable against density fluctuations (at which point it breaks down into individual clusters). Our computed value of $\theta_s$ is 0.0585$\pm$0.0010 \Am2. 
Analysis of our numerical results suggests that no metastable liquid phase exists in this system, at low $T$; that is, the system remains a solid all the way down to $\theta_s$, below which it breaks down into solid clusters. We also study the melting of the 2D triangular solid, and determine its melting temperature at approximately 6.8 K.

Although the PIMC method utilized in this work allows for the sampling of permutations of particles, which is essential in order to reproduce in the simulation any effect due to quantum statistics, permutations are not seen to occur, in the temperature range explored. This is because in the crystal phase, the only one observed here at low $T$, permutations are suppressed (as in most solids) by the localization of \paraH2 molecules; at higher temperature, on the other hand, though the crystal melts and molecules are less localized, they also behave more classically, as their thermal wave length decreases. Consistently with permutations not being important, i.e. \paraH2 molecules obeying Boltzmann statistics in the temperature range explored here, no evidence of SF can be seen. \cite{note} 

Our system of interest
is modeled as an ensemble of $N$ \paraH2 molecules, regarded as point particles and whose motion is restricted to two physical dimensions.
The quantum-mechanical many-body Hamiltonian is the following:
\begin{equation}\label{one}
\hat H = -{\hbar^2\over 2m}\sum_{i=1}^N \nabla_i^2 + \sum_{i<j} V(r_{ij}) 
\end{equation}
The system is enclosed in a simulation cell shaped as a parallelogram of area $A$, with periodic boundary conditions in all directions. The density is $\theta$=$N/A$. 
In Eq. (\ref {one}), $m$ is the \paraH2 molecular  mass and 
$V$ is the potential describing the interaction between two \paraH2  molecules,
only depending on their relative distance. 
The Silvera-Goldman potential\cite{silvera78} was chosen to model the interaction $V$, mostly for consistency with existing, comparable calculations. However, this potential has also been shown to provide an  acceptable quantitative description of bulk condensed \paraH2. \cite{johnson96,operetto04} 

The PIMC method is a numerical (Quantum Monte Carlo) technique that allows one to obtain accurate estimates of physical averages for quantum many-body systems at finite temperature. 
The only input of a PIMC calculation is  the many-body Hamiltonian (\ref{one}) (i.e., the potential energy function $V$). 
Because thorough descriptions of PIMC exist, \cite{ceperley95} it will not be reviewed here. The main technical details of this calculation are illustrated in Ref. \onlinecite{boninsegni04}.

Most of the results provided in this manuscript pertain to a system of $N$=64 molecules.
We also obtained results for a system of 144 particles at the $T=0$ equilibrium density $\theta_\circ$ and near the spinodal density $\theta_s$. At the beginning of the simulation, molecules are arranged on a triangular lattice. No significant dependence of the estimates on the size of the system can be observed for the physical quantities studied here, with the exception of the energy (see below).
\begin{figure}[h]
\centerline{\includegraphics[height=4in,angle=-90]{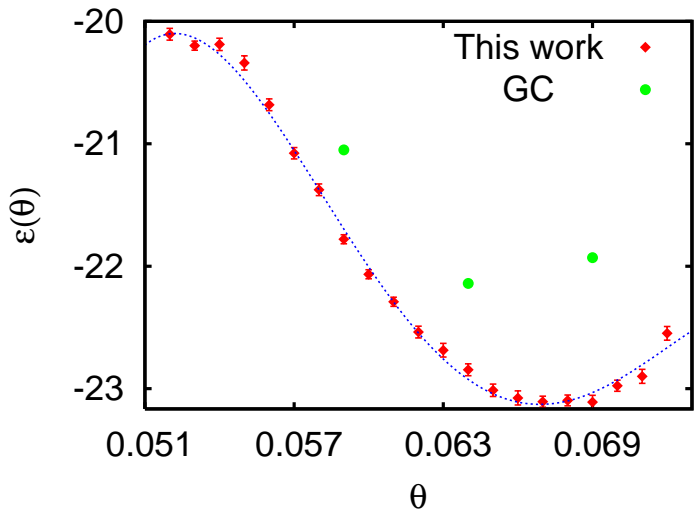}}
\caption{Energy per  molecule  $\epsilon$ (in K)  vs. coverage $\theta$  (\AA$^{-2}$), computed by PIMC for a 2D \paraH2 film of 64 molecules (diamonds). Estimates shown are obtained at a temperature $T$=2 K, but are indistinguishable, within statistical uncertainties, from those computed at lower temperatures. Dashed line is a polynomial fit to the data. Also shown for comparison (circles) are the results of the same calculation carried out by Gordillo and Ceperley (Ref. \onlinecite{gordillo97}), also using PIMC (at T=1 K) but on a system of  30 molecules or less. } \label{fig1}
\end{figure}

Fig. \ref{fig1} shows computed values of the energy per \paraH2 molecule (in K) vs the 2D density (coverage) $\theta$, expressed in \Am2, for a system of $N$=64 molecules. Energy estimates are found to be nearly independent on temperature, below $T$ $\approx$ 3 K. Thus, the results shown in Fig. \ref{fig1} are essentially ground state estimates. A polynomial fit of the data yields an equilibrium density     
$\theta_\circ$ (corresponding to the minimum of the $\epsilon(\theta)$ curve) of 0.0668$\pm$0.0005 \Am2. 
This result is in agreement with a recent, independent  Quantum Monte Carlo calculation at zero temperature. \cite{cazorla04}  Also shown in Fig. \ref{fig1} are the estimates obtained by Gordillo and Ceperley, who carried out PIMC calculations at low temperature on a system of thirty molecules or less. The two calculations agree, insofar as locating the equilibrium density; there is a numerical discrepancy between our calculation and theirs, which can be attributed to the difficulty of determining quantitatively, on a small size system,  the contribution to the potential energy associated to the periodic images of the system outside the simulation cell. For, the  H$_2$ intermolecular potential has a long-range attractive tail whose overall contribution to the potential energy is considerably greater than that obtained, for example, for condensed Helium at equilibrium, in spite of the fact that the interparticle potentials decay as 1/$r^6$ at long distance in both cases. All of our energy estimates are obtained by computing the above-mentioned contribution to the potential energy by setting the value of the pair correlation function to one outside the simulation  cell. On comparing results obtained on systems with $N$=64 and $N$=144 particles, we estimate the systematic error on the energy values furnished here, due to the finite size of the system, to be less than 0.15 K per particle when $N$=64, and less than 0.03 K per particle for $N$=144.
\begin{figure}
\centerline{\includegraphics[height=4in,angle=-90]{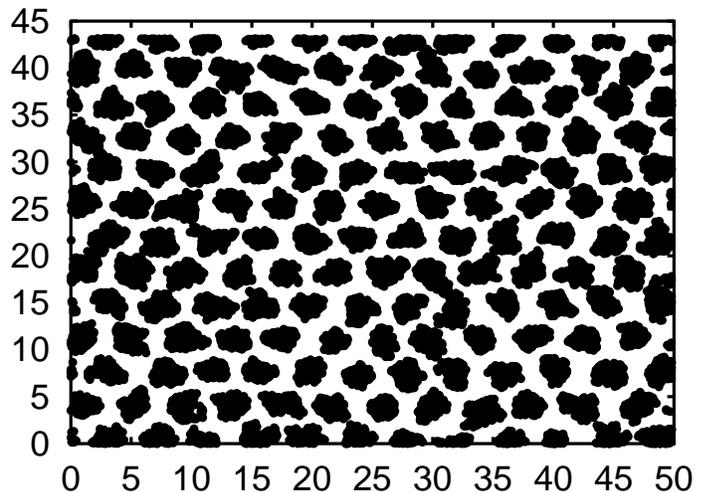}}
\caption{Typical many-particle configuration generated by the PIMC simulation at $T$=2 K, for a system of 144 \paraH2 molecules at a density $\theta=0.067$ \Am2, i.e., near the computed equilibrium density. Periodic boundary conditions are used in both directions. The arrangement of molecules on a triangular lattice is clearly seen. Each black ``cloud" consists of 320 points, each representing the position of a \paraH2 molecule along its path in imaginary time (see, for instance, Ref. \onlinecite{ceperley95}). } 
\label{fig2}
\end{figure}

Fig. \ref{fig2} shows a typical many-particle configuration, generated by the PIMC simulation, for a system of 144 \paraH2 molecules at a density $\theta$=0.067 \Am2 (i.e., close to $\theta_\circ$), at a temperature $T$=2 K. The arrangement of molecules on a triangular lattice is clearly seen. Each fuzzy ``cloud" represents a \paraH2 molecule, the typical size of each cloud being a measure of quantum delocalization. There is essentially no overlap of clouds associated to different molecules, which is qualitatively an indication that quantum exchanges are unimportant, and molecules can be regarded as obeying essentially Boltzmann statistics.

\begin{figure}
\centerline{\includegraphics[height=4in,angle=-90]{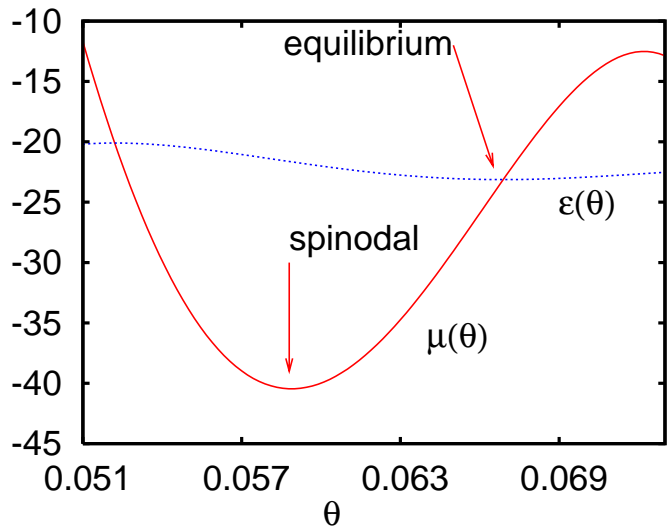}}
\caption{Chemical potential $\mu(\theta)$ (in K) versus density $\theta$ (\Am2) computed based on Eq. (\ref{mu}) for a 2D \paraH2 system, using the low temperature energy data shown in Fig. \ref{fig1}. Also shown (dotted line) is the polynomial fit of the $\epsilon(\theta)$ results of Fig. \ref{fig1}. } 
\label{fig3}
\end{figure}

Using the polynomial fit to the energy data shown in Fig. \ref{fig1}, we computed the low temperature chemical potential $\mu(\theta)$ through
\begin{equation}\label{mu}
\mu(\theta)=\epsilon(\theta)+\theta\frac{d\epsilon}{d\theta}
\end{equation}
The chemical potential is shown in Fig. \ref{fig3} (solid line). The equilibrium density $\theta_\circ$ is identified by the condition $\mu(\theta_\circ)=\epsilon(\theta_\circ)$. A second density of interest is the {\it spinodal} ($\theta_s$), corresponding to the condition $d\mu/d\theta=0$.  Based on our $\epsilon(\theta)$ data, we obtain $\theta_s=0.0585\pm0.0010$ \Am2. 

\begin{figure}
\centerline{\includegraphics[height=3.81in,angle=-90]{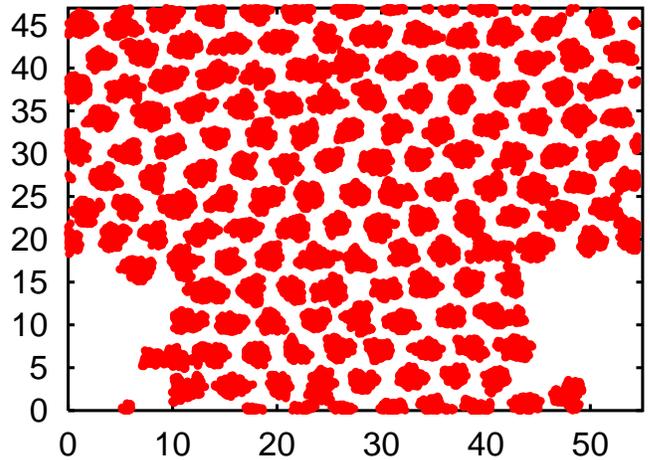}}
\caption{Same as in Fig. \ref{fig2} but  at a density $\theta=0.056$ \Am2, i.e., below the spinodal density $\theta_s$ (see text). } 
\label{fig2a}
\end{figure}

The spinodal density is the lowest density down to which the uniform phase can be {\it stretched} (at negative pressure), before becoming unstable against density fluctuations. At any density lower than $\theta_s$, the uniform film breaks down into ``puddles".  We can observe this effect directly, by examining many-particle configurations generated by a PIMC simulation of a system of 144 \paraH2 molecules; an example is shown in Fig. \ref{fig2a}, for 2D \paraH2 at a density $\theta$=0.056 \Am2, i.e., slightly lower than $\theta_s$. Such visual observations strongly suggest that such puddles are not liquid; rather, the system retains therein its triangular crystal structure.

In their 1997 calculation, Gordillo and Ceperley made the suggestion that a metastable liquid phase may exist at densities below $\approx$ 0.059 \Am2; however, they did not attempt to locate the spinodal density in their study. \cite {gordillo97} As it turns out, $\theta_s$ as obtained in this work lies precisely in correspondence of  their proposed location of the solid-liquid transition.
By direct observation of configurations such as those shown in Figs. \ref{fig2} and \ref{fig2a}, as well as by examining the pair correlation function and the static structure factor, we have confirmed the conclusion of Ref. \onlinecite {gordillo97}, namely that the system is a crystal, for $\theta > \theta_s$.  Because the uniform phase breaks down below $\theta_s$, 
we further conclude that no metastable liquid phase of \paraH2 exists in 2D, in the $T$$\to$0 limit. 
The system remains a solid all the way down to $\theta_s$, and breaks down into solid clusters at lower densities.
\begin{figure}
\centerline{\includegraphics[height=3.81in,angle=-90]{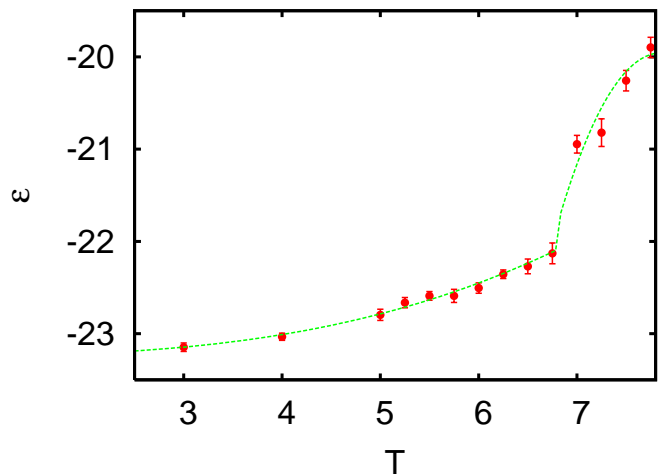}}
\caption{Energy $\epsilon$ (K) per \paraH2 molecule as a function of temperature (K), at $\theta=0.067$ \Am2. Numerical estimates pertain to a system of $N$=144 \paraH2 molecules. The extrapolated $T$=0 value is -23.25$\pm$0.05 K. The clear change of slope at $T$$\approx$ 6.8 K corresponds to the melting of the 2D 
triangular lattice. Dashed line is a guide to the eye.} 
\label{fig3a}
\end{figure}
In order to study the melting of the equilibrium 2D crystal, we computed the energy per particle $\epsilon(T)$ as a function of temperature for a system of $N$=144 particles. For simplicity, we have kept the density fixed at its $T$=0 equilibrium value, namely 0.067 \Am2. The results are shown in Fig. \ref{fig3a}. 

At low $T$, $\epsilon(T)$ follows the expected $\sim$ $T^3$ behavior, which is consistent with phonons being the low-lying excitations of the 2D quantum crystal. The extrapolated $T$=0 energy value is -23.25$\pm$0.05 K, which is in satisfactory agreement with the value of -23.4 K recently reported by Cazorla and Boronat, who carried out a $T$=0 calculation. We estimate our uncertainty in the determination of the potential energy to amount to less than 0.03 K per molecule, on a system of this size. There is a difference of approximately 0.1 K between the above, extrapolated energy value, and that obtained on a system of 64 particles, the latter being the higher one.

A sudden change of slope is seen to occur at $T_m$=6.8 K, where the specific heat $c(T)=d\epsilon/dT$ takes on a peak, which corresponds to the melting of the 2D crystal. This value of the melting temperature is comparable to that found in studies of \paraH2 films on alkali metal substrates \cite{boninsegni04}, and slightly higher than that of a \paraH2 surface (computed by PIMC), \cite{wagner96} and of an adsorbed \paraH2 monolayer .\cite {vilches92} This suggests that melting may occur at a lower temperature in three dimensions, as zero-point motion of molecules in the direction perpendicular to the substrate promotes evaporation. In any case, these melting temperatures are still too high to expect that a superfluid transition of \paraH2 may be observed.

In conclusion, we have carried out extensive PIMC studies of 2D condensed \paraH2, determining its low temperature equation of state, inferring its equilibrium and spinodal densities, as well as the melting temperature of the equilibrium system. Consistently with previous study, we found the system to be a triangular crystal at equilibrium; this 2D crystal melts at a temperature of approximately 6.8 K. 
 
We have found no evidence of any metastable liquid phase at low $T$. The system is found to remain a solid all the way down to the spinodal density, below which it breaks down into solid clusters. This result, perhaps unexpected, raises some doubts about the long term prospects of observing a superfluid phase 
of \paraH2 upon supercooling the  liquid.
Observing novel phases of \paraH2, including a (superfluid) liquid, may 
require achieving a  substantial renormalization of the interaction of 
\paraH2 molecules, possibly through their interaction with the surface 
electrons of a metal substrate, or of a nanostructure. An alternate route might be that suggested in Ref. \onlinecite {gordillo97}, namely stabilizing a liquid at low temperature by means of an external potential incommensurate with the crystal structure of \paraH2.

This work was supported in part by the Petroleum Research Fund of the American 
Chemical Society under research grant 36658-AC5, and by the Natural Science 
and Engineering Research Council of Canada under research grant G121210893. 


\begin{thebibliography}{99}
\bibitem{ginzburg72}
V. L. Ginzburg and A. A. Sobyanin,  JETP Letters {\bf 15}, 242 (1972).
\bibitem{bretz81}
M. Bretz and A. L. Thomson,
Phys. Rev. B {\bf 24}, 467 (1981).

\bibitem{maris86}
G. M. Seidel, H. J. Maris, F. I. B. Williams and J. G. Cardon,
Phys. Rev. Lett. {\bf 56}, 2380 (1986).

\bibitem{maris87}
H. J. Maris, G. M. Seidel and F. I. B. Williams, 
Phys. Rev. B {\bf 36}, 6799 
(1987).

\bibitem{schindler96}
M. Schindler, A. Dertinger, Y. Kondo and F.  Pobell, 
Phys. Rev. B {\bf 53}, 11451 (1996).


\bibitem{nielsen80}
M. Nielsen, J. P. McTague and L. Passell in {\it Phase Transitions in Surface Films}, edited by J. Dash and J. Ruvalds (Plenum, New York, 1980).

\bibitem{lauter90}
H. J. Lauter, H. Godfrin, V. L. P. Frank and P. Leiderer in {\it Phase Transitions in Surface Films 2}, edited by H. Taub, G. Torzo, H. J. Lauter and S. C. Fain Jr. (Plenum, New York, 1990).

\bibitem{wiechert91}
H. Wiechert in {\it Excitations in Two-Dimensional and Three-Dimensional Quantum Fluids}, edited by A. F. G. Wyatt and H. J. Lauter (Plenum, New York, 1991).

\bibitem{vilches92}
F. C. Liu, Y. M. Liu and O. E. Vilches, J. Low Temp. Phys. {\bf 89}, 649 (1992).
\bibitem{boronat00}
M. C. Gordillo, J. Boronat and J. Casulleras, Phys. Rev. Lett. {\bf 85}, 02348 (2000).
\bibitem{gordillo97}
M. C. Gordillo and D. M. Ceperley, Phys. Rev. Lett. {\bf 79}, 3010 (1997).

\bibitem{wagner94}
M. Wagner and D. M. Ceperley, J. Low Temp. Phys. {\bf 94}, 161 (1994).

\bibitem{wagner96}
M. Wagner and D. M. Ceperley, J. Low Temp. Phys. {\bf 102}, 275 (1996).

\bibitem{nho02}
K. Nho and E. Manousakis, { Phys. Rev. B} {\bf 65}, 115409 (2002).
\bibitem{nho03}
K. Nho and E. Manousakis, Phys. Rev. B {\bf 67}, 195411 (2003).
\bibitem{shi03}
W. Shi, J. K. Johnson and M. W. Cole, Phys. Rev. B {\bf 68}, 125401 (2003).
\bibitem{boninsegni04}
M. Boninsegni, Phys. Rev. B {\bf 70}, 125405 (2004).
\bibitem{note}
We have not attempted to search for a possible superfluid phase of {\it solid} \paraH2. The
absence of many-particle permutations observed in this work leads us to believe that, if such a phase exists, it must be at significantly lower temperatures than those considered here.
\bibitem{silvera78}
I. F. Silvera and V. V. Goldman, {J. Chem. Phys.} {\bf 69}, 4209 (1978).
\bibitem{johnson96} Q. Wang, J. K. Johnson and J. Q. Broughton, {Mol. Phys.} {\bf 89}, 1105 (1996).
\bibitem{operetto04} F. Operetto and F. Pederiva, Phys. Rev. B {\bf 69}, 024203 (2004).
\bibitem{ceperley95}
D. M. Ceperley, {Rev. Mod. Phys.} {\bf 67}, 279 (1995).

\bibitem{voth}
S. Jang, S. Jang and G. A. Voth, J. Chem. Phys. {\bf 115}, 7832 (2001).
\bibitem{pollock84}
E. L. Pollock and D. M. Ceperley, Phys. Rev. B {\bf 30}, 2555 (1984).
\bibitem{cazorla04}
C. Cazorla and J. Boronat, J. Low Temp. Phys. {\bf 134}, 43 (2004).
\end{thebibliography}
\end{document}